\documentclass[twocolumn,showpacs,prl]{revtex4}

\usepackage{amssymb,amsmath} 
\usepackage{bbm} 
\usepackage{bm} 
\usepackage[english]{babel}

\usepackage{graphicx}
\usepackage{color}

\usepackage{float}

\usepackage[bookmarks,breaklinks,plainpages=false,pdfpagelabels]{hyperref}  
	\hypersetup{linkcolor=red,citecolor=blue,filecolor=dullmagenta,urlcolor=darkblue} 

\definecolor{grey}{rgb}{0.4,0.4,0.4}
\definecolor{dullmagenta}{rgb}{0.4,0,0.4}
\definecolor{darkblue}{rgb}{0,0,0.4}
\definecolor{orange}{rgb}{1,0.5,0}
\definecolor{lightbrown}{rgb}{0.75,0.5,0.25}
\definecolor{tan}{cmyk}{0.14,0.42,0.56,0}
\definecolor{djunglegreen}{cmyk}{0.99,0,0.52,0}
\definecolor{lightgreen}{rgb}{0,1,0}
\definecolor{olivegreen}{cmyk}{0.64,0,0.95,0.40}
\definecolor{midgreen}{rgb}{0.0,0.675,0.0}

\newcommand{\normsingle}[1]{\left|{#1}\right|}

\newcommand{\e}{\ensuremath{\epsilon}}

\newcommand{\s}{\ensuremath{\sigma}}

\newcommand{\q}{\quad}

\renewcommand{\.}{\hspace{0.3mm}}
\newcommand{\ms}{\hspace{-0.2mm}}

\newcommand{\ra}{\ensuremath{\rightarrow}}

\newcommand{\Grm}{\ensuremath{\mathrm{G}}}

\newcommand{\Rrm}{\ensuremath{\mathrm{R}}}

\newcommand{\Vrm}{\ensuremath{\mathrm{V}}}
\newcommand{\Wrm}{\ensuremath{\mathrm{W}}}

\newcommand{\arm}{\ensuremath{\mathrm{a}}}

\newcommand{\hrm}{\ensuremath{\mathrm{h}}}
\newcommand{\irm}{\ensuremath{\mathrm{i}}}

\newcommand{\prm}{\ensuremath{\mathrm{p}}}

\newcommand{\Ocal}{\ensuremath{\mathcal{O}}}
\newcommand{\Pcal}{\ensuremath{\mathcal{P}}}

\newcommand{\Scal}{\ensuremath{\mathcal{S}}}

\newcommand{\Zcal}{\ensuremath{\mathcal{Z}}}

\newcommand{\Jbbm}{\ensuremath{\mathbbm{J}}}

\newcommand{\onebbm}{\ensuremath{\mathbbm{1}}}

\newcommand{\kbm}{\ensuremath{\bm{k}}}

\newcommand{\pbm}{\ensuremath{\bm{p}}}

\newcommand{\xbm}{\ensuremath{\bm{x}}}
\newcommand{\ybm}{\ensuremath{\bm{y}}}

\newcommand{\D}{\ensuremath{\mathcal{D}}}
\renewcommand{\d}{\ensuremath{\mathrm{d}}}
\newcommand{\ee}{\ensuremath{\mathrm{e}}}

\newcommand{\defas}{\mathrel{\mathop :}=} 
\newcommand{\hph}[1]{\hphantom{#1\;\,}}
\newcommand{\Tr}{\ensuremath{\mathrm{Tr}}}

\renewcommand{\t}{\text}

\renewcommand{\ol}{\overline}

\newcommand{\hc}{{\rm h.c.}}
\newcommand{\rhs}{r.h.s.}

\newcommand{\wrt}{w.r.t.}
\newcommand{\cds}{\,\cdot\,}

\begin{document}

\title{Stochastic Inflation and Dimensional Reduction}

\author{Florian K{\"u}hnel}
\email{kuehnel@physik.uni-bielefeld.de}
\author{Dominik J. Schwarz}
\email{dschwarz@physik.uni-bielefeld.de}
\affiliation{Fakult{\"a}t f{\"u}r Physik, Universit{\"a}t Bielefeld, Postfach 100131, 33501 Bielefeld, Germany}
\date{\today}

\begin{abstract}
We adopt methods that are well known in statistical physics to the problem of stochastic inflation. The effective power spectrum for the classical, stochastic long-wavelength fluctuations is calculated for free scalar fields in a de Sitter background. For a smooth separation into long and short wavelengths, we identify an infra-red divergence of the effective power spectrum, which has its correspondence in statistical physics in the phenomenon of dimensional reduction. The inflationary dynamics pushes the affected scales exponentially fast to large superhorizon scales, and establishes scale-invariant behavior for smaller scales (for massless fields). In the limit of a sharp separation of wavelengths, the scale of the infra-red divergence is pushed to infinity.
\end{abstract}


\maketitle

The concept of {stochastic inflation}, introduced by {Starobinsky} several years ago \cite{1982PhLB..117..175S}, provides a framework to study the evolution of quantum fields in an inflationary universe \cite{1986LNP...246..107S,PhysRevD.27.2848} and has acquired considerable interest over the last years \cite{liguori-2004-0408,Tsamis:2005hd}. The key idea lies in splitting the quantum fields into long- and short-wavelength modes, and viewing the former as classical objects evolving in an environment provided by quantum fluctuations of shorter wavelengths. It constitutes an example of how the fundamental properties of quantum fields can be modeled using methods of statistical mechanics. From this point of view, one focusses on the ``relevant'' degrees of freedom, the long-wavelength modes, and regards the short-wavelength modes as ``irrelevant'' ones, generating a bath in which the former evolve. The natural length scale of this problem is the Hubble length, from which ``short'' and ``long'' acquire their physical meaning.

This stochastic description of cosmological inflation has in fact two steps of complexity. In a simplified setup, the problem is reduced to describe a scalar test field $\varphi$ in a fixed cosmological background. If $\varphi$ is free, massive and minimally coupled, one obtains after splitting into long and short wavelengths, $\varphi = \varphi_{\t{\tiny L}} + \varphi_{\t{\tiny S}}$, an effective equation of motion of generalised Langevin-type,
\begin{align}
	( \Box + \mu^{2} ) \varphi_{\t{\tiny L}}( t, \xbm )
		&=						\hrm( t, \xbm ) .
								\label{eq:linearfieldeq}
\end{align}
Here, $\varphi_{\t{\tiny L}}$ is viewed as a classical entity, evolving stochastically in the presence of a (quantum) random force $\hrm$, which is {Gaussian} distributed with zero mean.

Early studies focussed on homogeneous fields---thus restricting attention to the {time} evolution of $\varphi_{\t{\tiny L}}$. The study of inhomogeneous fields (see, e.g., \cite{liguori-2004-0408}) is more involved, but also allows one to discuss {spatial} correlations.

In the full problem of stochastic inflation, $\varphi$ is the inflaton field, rather than a test field in a fixed background. This requires not only a stochastic description of $\varphi$, but also of the geometry.  Thus, one should analogously split the space-time metric and curvature into long- and short-wavelength parts.

In this work, we study the scaling behavior and time evolution of the power spectrum of $\varphi_{\t{\tiny L}}$ in a fixed background. We do so by means of replica field theory, which is well known in statistical physics \cite{1991JPhy1...1..809M}. As far as we know, replica fields have not been applied to cosmology before. They allow us to compute spatial correlations and their time evolution for the general test-field case.

Below we illustrate this method by the example of a free, minimally coupled, $N$-component, real field $\vec{\varphi}$ with mass $\mu$. Let us further restrict our attention to a spatially-flat de Sitter universe in $d$ space-time dimensions. Its scale factor is $a(t) = \ee^{H t}$ with Hubble expansion rate $H$. For convenience we rescale to dimensionless variables $\vec{\varphi} / H^{(d - 2) / 2} \ra \vec{\varphi}$, $\mu / H \ra \mu$, $t H \ra t$, $\xbm H \ra \xbm$, and $\kbm / H \ra \kbm$, and use $\hslash = c = 1$. The mode function $u( t, k)$ is defined via the decomposition of the field components, $i = 1, \ldots, N$,
\begin{align}
	\varphi_{i}( t, \kbm )
		&=						\hat{\arm}_{i}( \kbm ) u( t, k ) + \hc,
								\label{eq:varphi-fourier-decomposition}
\end{align}
with $k \defas \normsingle{\kbm}$ and the annihilation and creation operators obey the commutation relations
\begin{align}
\begin{split}
	\big[ \hat{\arm}_{i}( \kbm ), \hat{\arm}_{j}^{\dagger}( \pbm ) \big]
		&=						( 2 \pi )^{d - 1} \delta^{d - 1}( \kbm - \pbm ) \,\delta_{i j},\\
	\big[ \hat{\arm}_{i}( \kbm ), \hat{\arm}_{j}( \pbm ) \big]
		&=						0.
\end{split}
\end{align}
The propagator is defined as
\begin{align}
	\Grm_{0}( t, t'\!, \kbm, \kbm' )
		&\defas					\frac{1}{N} \big\langle \Omega \big| \vec{\varphi}( t, \kbm) \cdot \vec{\varphi}( t'\!, \kbm' ) \big| \Omega \big\rangle ,
\end{align}
where the vacuum $| \Omega \rangle$ is defined by $\hat{\arm}( k ) | \Omega \rangle = 0$ at $t = 0$ and a subscript ``$0$'' indicates a quantity that is calculated in the absence of any noise.

An object of central interest in cosmology is the dimensionless power spectrum $\Pcal_{\!\varphi}( k )$. Its relation to some field propagator $\Grm\!\left( k \right)$ with an infra-red behavior $\Grm\!\left( k \right) \sim k^{-\eta}$, i.e., for $k \ll a H$, is given by
\begin{align}
	\Pcal_{\!\varphi}( k )
		&\defas					k^{d-1} \Grm\!\left( k \right)
		\sim						k^{n_{\varphi} - 1},
								\label{eq:spectral-index}
\end{align}
with the spectral index $n_{\varphi}$, connected to the critical exponent $\eta$ via $n_{\varphi} = d - \eta$. For $d = 4$ and $\mu = 0$ the power spectrum of the free, noiseless theory is scale-invariant, i.e., $n_{\varphi} = 1$ for $\eta_{0} = 3$. Non-zero mass leads to
\begin{align}
	n_{\varphi}
		&=						4 - 3 \sqrt{1 - \frac{ 4 }{ 9 } \mu^{2} }
		=						1 + \frac{ 2 }{ 3 } \mu^{2} + \Ocal\!\left( \mu^{4} \right) .
								\label{eq:ns}
\end{align}
Note that this result does not include any metric perturbation, which would cause the deviation from $n_{\varphi} = 1$ to be negative (see, e.g., \cite{PhysRevD.66.023515}).

Let us now introduce a split of the quantum field $\vec{\varphi}$ into short- and long-wavelength modes. For the free scalar field under consideration, this generates a linear noise term on the \rhs~of the field equation, which becomes of type \eqref{eq:linearfieldeq}. On the level of the action, this corresponds to a linear random potential
\begin{align}
	\Vrm\!\left( \.\vec{\varphi}_{\t{\tiny L}} \right)
		&=						 \vec{\hrm}\cdot\vec{\varphi}_{\t{\tiny L}}
								\label{eq:linear-random-potential}
\end{align}
plus a term quadratic in $\hrm$, which is ascribed to a Gaussian probability distribution $\prm[ \{ \hrm \} ]$. In statistical field theory, this is the so-called random-field case.

For interacting scalar fields, the split of $\vec{\varphi}$ into long and short wavelengths generates higher powers of both $\vec{\varphi}_{\t{\tiny L}}$ and the noise in the equation of motion. They may as well be collected into a probability distribution and a random potential, which might be written as
\begin{align}
	\Vrm\!\left( \.\vec{\varphi}_{\t{\tiny L}} \right)
		&=						-  \sum_{j=1}^{\infty}\sum_{\{i_{1}, \dots, i_{j}\} = 1}^{N} \hrm_{i_{1}\ldots i_{j}} 
								\varphi_{\t{\tiny L}}^{i_{1}} \ldots \varphi_{\t{\tiny L}}^{i_{j}},
								\label{eq:Vrmgeneral}
\end{align}
with the Taylor coefficients $\{ \hrm \}$ being a set of random variables subject to $\prm[ \{ \hrm \} ]$.

For any quantity $\Ocal$ depending on $\{ \hrm \}$, the stochastic average, which shall be denoted by a bar, is then calculated as
\begin{align}
	\ol{\Ocal[ \{ \hrm \} ]}	
		&\defas					\int{\!\D[ \{ \hrm \} ]}\,\prm[ \{ \hrm \} ]\,\Ocal[ \{ \hrm \} ]  .
\end{align}
Note, that linearizing the equation of motion in the quantum modes, corresponds {\it exactly} to a Gaussian distribution with vanishing mean. As has been shown recently, this effectively re-sums the leading-log contribution of the full quantum theory \cite{Tsamis:2005hd}.

Objects of central interest are $n$-point correlation functions of the classical long-wavelength field $\vec{\varphi}_{\t{\tiny L}}$, i.e.,
\begin{align}
	\overline{\big\langle \varphi_{\t{\tiny L}}^{i_{1}}( x_{1} ) \cdot \ldots \cdot \varphi_{\t{\tiny L}}^{i_{n}}( x_{n} ) \big\rangle} .
\end{align}
Since they can be derived from the generating functional
\begin{align}
	\Zcal[ \.\vec{\jmath}_{\t{\tiny L}} ]
		&=						\int{\D[ \.\vec{\varphi}_{\t{\tiny L}} ]}\exp{ \big\{ \irm\. \Scal[ \.\vec{\varphi}_{\t{\tiny L}}, \vec{\jmath}_{\t{\tiny L}} \,] \big\} }
								\label{eq:Z}
\end{align}
by taking derivatives of $\ln\{ \Zcal[ \.\vec{\jmath}_{\t{\tiny L}} \,] \}$ \wrt~an external source $\vec{\jmath}_{\t{\tiny L}}$, one in general needs to know $\overline{\ln\{ \Zcal[ \.\vec{\jmath}_{\t{\tiny L}} \,] \}}$. This might be difficult to calculate, because one has to average a logarithm of a path integral over an exponential. To perform these stochastic averages we use the replica trick \cite{1988PhT....41l.109M},
\begin{align}
	\overline{ \ln\{ \Zcal \} }
		&=						\lim_{m \rightarrow 0} \frac{ 1 }{ m } \ln\big\{ \overline{ \Zcal^{m} } \big\} .
\end{align}
Thus, one just has to compute $\overline{ \Zcal^{m} }$ for {integer} $m$, and, if the result is analytic in $m$, to take $m \ra 0$ at the end. This tantamounts to the introduction of $m$ different copies (replicas) of the same system which are then coupled through the noise average.

As already mentioned, the noise distribution $\prm[ \{ \hrm \} ]$ is Gaussian for free scalar fields, with the first and second cumulants given by
\begin{align}
\begin{split}
	\overline{\hrm_{i}( x )}
		&=						0 ,\\
	\overline{\hrm_{i}( x )\,\hrm_{j}( y )}
		&=						\delta_{ij}\,\phi( x, y ) ,
\end{split}
\end{align}
with a known function $\phi( x, y )$ depending on derivatives of the mode functions in \eqref{eq:varphi-fourier-decomposition}. In terms of the linear random potential \eqref{eq:linear-random-potential}, this may be written as
\begin{align}
\begin{split}
	\overline{\Vrm\ms \big( \.\vec{\varphi}_{\t{\tiny L}}^{\,a}( x ) \big)}
		&=						0 ,\\
	\overline{\Vrm\ms \big( \.\vec{\varphi}_{\t{\tiny L}}^{\,a}( x ) \big)\,\Vrm\ms \big( \.\vec{\varphi}_{\t{\tiny L}}^{\,b}( x ) \big)}
		&=						\phi( x, y )\,\vec{\varphi}_{\t{\tiny L}}^{\,a}( x ) \cdot \vec{\varphi}_{\t{\tiny L}}^{\,b}( y ) ,
									\label{eq:Vbar-cummulants}
\end{split}
\end{align}
where the replica indices $a, b = 1, \ldots, m$ label the different copies arising by means of the replica trick.

For interacting scalar fields, a Gaussian distribution is an approximation, but for many cases at least a reasonable staring point. Hence, one may specify a general random potential $\Vrm$ as in \eqref{eq:Vrmgeneral}, in analogy to the case of free scalar fields,
\begin{align}
	\overline{\Vrm\ms \big( \.\vec{\varphi}_{\t{\tiny L}}^{\,a}( x ) \big) \.\Vrm\ms \big( \.\vec{\varphi}_{\t{\tiny L}}^{\,b}( y ) \big)}
		&=						\phi( x, y ) N 
								\Rrm\!\left( \frac{ \vec{\varphi}_{\t{\tiny L}}^{\,a}( x ) \cdot \vec{\varphi}_{\t{\tiny L}}^{\,b}( y ) }{ N } \right) ,
								\label{eq:vvbar}
\end{align}
where each of the random variables $\{ \hrm \}$ is taken to be {Gaussian}-distributed with mean zero. Of course, the concrete form of the function $\Rrm$, as well as the form of its argument has to be determined from first principles. Note that for our example of a free scalar field, the Gaussian distribution is exact and that the function $\Rrm$ is linear, as seen in \eqref{eq:Vbar-cummulants}. In the terminology of statistical field theory, the space-time correlation $\phi( x, y )$ is called short-range, if $\phi( x, y ) = \wp( t, t' ) \delta^{d - 1}( \xbm - \ybm )$, and long-range for all other cases (c.f.~\cite{fedorenko-2007}). The latter is relevant for stochastic inflation.

After averaging over the noise we obtain from $\overline{ \Zcal^{m}}$ and equation \eqref{eq:vvbar} the replicated action
\begin{align}
	&{\Scal}^{(m)}
		=						\frac{ 1 }{ 2 }\sum_{a = 1}^{m}\int_{t, t'}\int_{\kbm}{{{\Grm_{\t{\tiny L}}}}_{0}^{-1}}( t, t'\!, k )\,
								\vec{\varphi}_{\t{\tiny L}}^{\,a}( t, k )\cdot\vec{\varphi}_{\t{\tiny L}}^{\,a}( t'\!,-k )
								\notag\displaybreak[1]\\
		&\q\hph{=}					- \frac{ 1 }{ 2 }\sum_{a,b = 1}^{m}\int_{x,y}\phi( x, y ) N \Rrm\!\left( \frac{ \vec{\varphi}_{\t{\tiny L}}^{\,a}( x ) \cdot \vec{\varphi}_{\t{\tiny L}}^{\,b}( y ) }{ N } \right),
								\label{eq:hn}
\end{align}
with the definitions $\int_{\kbm} \defas \int\d^{d-1}k / ( 2 \pi )^{d-1}$, $\int_{t} \defas \int\d t$, $\int_{x} \defas \int\d^{d}x$, and ${\Grm_{\t{\tiny L}}}_{0}\!\left( k \right) \defas \int_{\kbm'}{\Grm_{\t{\tiny L}}}_{0}\!\left( \kbm, \kbm' \right)$. We dropped the source term $\propto \vec{\jmath}_{\t{\tiny L}}$, because it is irrelevant for the following calculation. We see that the $m$ replicas are coupled to each other through the noise correlation \eqref{eq:vvbar}.

In the following, we will use a variational method to approximate action \eqref{eq:hn} by the {Gaussian} variational action
\begin{align}
	\Scal_{0}^{(m)}
		&\defas					\frac{ 1 }{ 2 }\sum_{a,b = 1}^{m}\int_{t, t'} \int_{\kbm}{{{\Grm_{\t{\tiny L}}}}^{-1}}_{ab}( t, t'\!, k )\,\vec{\varphi}_{a}( t, k )\cdot\vec{\varphi}_{b}( t'\!,-k )
								\label{eq:variational-action}
\end{align}
with the inverse propagator
\begin{align}
	{{\Grm_{\t{\tiny L}}}^{-1}}_{ab}
		\defas					{\Grm_{\t{\tiny L}}}_{0}^{-1} \delta_{ab} - \s_{ab}.
								\label{eq:Gab-inverse-ansatz}
\end{align}
Although Gaussian, the perhaps highly non-linear noise interaction $\Rrm( ..._{ab} )$ is accommodated by the non-diagonal replica structure $\s_{ab}$.

Let us comment on the structure of \eqref{eq:Gab-inverse-ansatz}. On the diagonal (in replica space) we find the inverse of the noiseless propagator ${\Grm_{\t{\tiny L}}}_{0}^{-1}$ plus some mass correction, to be determined later. This correction alone would not only be trivial but also inconsistent, as we will see later. Hence, the off-diagonal part is filled by some, a priori unknown, replica structure $\s_{ab}$, which, in general, can be time dependent, and, if one includes long-range noise correlation, also momentum dependent, directly affecting the scaling behavior of the power spectrum. Thus, although this variational method only generates a self-energy contribution, its off-diagonal replica structure might have a viable influence on large-scale correlations.

The Gaussian variational method becomes exact in the limit $N \ra \infty$ and allows one to go beyond ordinary perturbation theory. It is based on the following Feynman-Jensen inequality \cite{PhysRev.97.660}
\begin{align}
	\ln\{ \Zcal \}
		&\ge						\ln\{ \Zcal_{0} \}
								+ \Big\langle \Scal^{(m)}_{0} - \Scal^{(m)} \Big\rangle^{}_{\!0} ,
								\label{eq:fvar}
\end{align}
where the subscript $0$ refers to the variational action \eqref{eq:variational-action} and we temporarily Wickrotate to Euclidean signature. Equation \eqref{eq:fvar} can easily be proven by using the Jensen inequality $\exp\!\left\{ \langle \ldots \rangle \right\} \le \langle \exp\{ \ldots\} \rangle$ \cite{Jensen-1906-1}, which comes from the convexity of the exponential. The problem is to find the best ${\Grm_{\t{\tiny L}}}_{ab}$, i.e., the best $\s_{ab}$, satisfying (\ref{eq:fvar}) by maximizing the \rhs~of \eqref{eq:fvar}.

The result of the variation, again for Minkowski signature, is
\begin{align}
	\s_{ab}( t, p )
		&=						\int_{\xbm}\phi( t, \xbm ) \ee^{-\irm\pbm\cdot\xbm}\,
								\widehat{\Rrm}'\!\left( \int_{\kbm} \ee^{-\irm\kbm\cdot\xbm}\,{\Grm_{\t{\tiny L}}}_{ab}( t, k ) \right) ,
								\label{eq:sc+sab1N-times-lr}
\end{align}
where $\phi( t, \xbm ) \defas \phi( t, t, \xbm )$, $\s_{ab}( t, k ) \defas \s_{ab}( t, t, k )$ and ${\Grm_{\t{\tiny L}}}_{ab}( t, k ) \defas {\Grm_{\t{\tiny L}}}_{ab}( t, t, k )$, the prime denotes a derivative \wrt~the argument, and we assume translational invariance for the spatial correlation, i.e., $\phi( x, y ) = \phi( t, t'\!, \normsingle{\xbm - \ybm} )$. The function $\widehat{\Rrm}$ is defined by $\widehat{\Rrm}\big( \langle \;\cdot\; \rangle^{}_{0} \big) \defas \langle \Rrm( \;\cdot\; ) \rangle^{}_{0}$. In the limit $N \ra \infty$, or trivially for free scalar fields for arbitrary values of $N$, one has $\widehat{\Rrm} = \Rrm$.

The physical interpretation of the saddle-point equations \eqref{eq:Gab-inverse-ansatz} and \eqref{eq:sc+sab1N-times-lr} is the following: The replica structure $\s$ is a generalized self-energy (c.f. the discussion in section 3 of \cite{1991JPhy1...1..809M}).

The replica structure \eqref{eq:sc+sab1N-times-lr} already contains our main result, namely the dimensional reduction at large scales, as will become clear in the remainder of this work. It also shows that the replica matrix is in general space-time dependent, which affects the scaling behavior of the two-point function (c.f.~\cite{PhysRevB.27.5875}). As can easily be seen, in this (replica) Gaussian variational approximation, {any} interaction without random variables is diagonal in replica space. Therefore it only modifies the diagonal part of the self-energy. Arbitrary noise gives rise to a complex replica structure, however.

For the free scalar field studied in this paper, one observes that random-field noise (linear potential) {\it necessarily} yields a uniform replica matrix, i.e., $\s_{ab} \equiv \s$ for all $a,b$, because $\widehat{\Rrm}' = \t{const.}$ This is the so-called replica symmetric case, which has intensively been studied \cite{1991JPhy1...1..809M}. It means that different replicas couple all in the same way among each other. To leading order in the number of replicas, $m$, we find that $( {\Grm_{\t{\tiny L}}}_{ab} )$ has the form
\begin{align}
	( {\Grm_{\t{\tiny L}}}_{ab} )( t, k )
		&=						{\Grm_{\t{\tiny L}}}_{0}( t, k ) \onebbm + \s( t, k )\,{\Grm_{\t{\tiny L}}}_{0}( t, k )^{2} \Jbbm .
								\label{eq:grs}
\end{align}
where ${\Grm_{\t{\tiny L}}}_{0}( t, k ) \defas {\Grm_{\t{\tiny L}}}_{0}( t, t, k )$, and the $m \times m$-matrix $\Jbbm$ has $1$ in every entry. In the limit of vanishing correlation, i.e., $\s \ra 0$, the free theory is recovered. To derive \eqref{eq:grs}, no assumption except the validity of the Gaussian variational method is made.

The {physical} propagator ${\Grm_{\t{\tiny L}}}$ of the long-wavelength field $\vec{\varphi}_{\t{\tiny L}}$ is obtained from ${\Grm_{\t{\tiny L}}}_{ab}$ via
\begin{align}
	{\Grm_{\t{\tiny L}}}( t, k )
		&=						\lim_{m \ra 0} \frac{ 1 }{ m } \Tr\big[ ( {\Grm_{\t{\tiny L}}}_{ab} )( t, k ) \big] ,
								\label{eq:armean}
\end{align}
which is simply the arithmetic mean of the trace of the replica matrix propagator (c.f. section 4 of \cite{1991JPhy1...1..809M}) and yields
\begin{align}
	\Grm_{\t{\tiny L}}( t, k )
		&=						{\Grm_{\t{\tiny L}}}_{0}( t, k ) + \s(t, k)\,{\Grm_{\t{\tiny L}}}_{0}( t, k )^{2} .
								\label{eq:G=G0-sigmaG02}
\end{align}

We now turn to the study of the infra-red behavior of the physical propagator and therefore the power spectrum, with focus on the spectral index. Although irrelevant for stochastic inflation, we first assume short-range correlation for pedagogical reasons. Equation \eqref{eq:sc+sab1N-times-lr} then implies $\s$ to be independent of momentum. With the infra-red behavior ${\Grm_{\t{\tiny L}}}_{0}( t, k ) \sim k^{-\eta_{0}}$ and $\Grm_{\t{\tiny L}}( t, k ) \sim k^{-\eta}$ ($\eta_{0}, \eta > 0$), we obtain with the help of \eqref{eq:grs} and \eqref{eq:G=G0-sigmaG02} the relation
\begin{align}
	\eta
		&=						2 \eta_{0}
								\label{eq:eta-DR-RS}
\end{align}
---drastically different from to the noiseless case, where $\eta = \eta_{0}$. This is a variant of the phenomenon of dimensional reduction \cite{PhysRevLett.43.744,1984CMaPh..94..459K}. It is best understood in $\xbm$-space, where the two-point function goes in the infra-red like $\normsingle{ \xbm }^{- ( d - 1 ) + \eta}$, and the change in the exponent $\eta$ can be absorbed into the {reduced} dimension $d' \defas d - \eta_{0}$. We have to stress that, in the random-field case and for short-range correlation, the dimensional reduction theorem can rigorously be proven to hold in great generality, i.e., to all orders in perturbation theory and for arbitrary non-random potentials (see especially \cite{1984CMaPh..94..459K} for a supersymmetric version of the proof).

Let us now consider the case of long-range correlation of the form $\phi( \xbm ) \sim \normsingle{\xbm}^{- ( d - 1 ) + \rho}$ with $\rho < d - 1$. This describes properly the infra-red limit of the physical model discussed below. In momentum space, the above choice implies $\phi( k ) \sim k^{- \rho}$ and hence $\s_{ab}( k ) \sim k^{- \rho}$ by virtue of \eqref{eq:sc+sab1N-times-lr}. For $\rho > - \eta_{0}$, the infra-red behavior of the power spectrum deviates from the noiseless result, and we find $\eta = 2 \eta_{0} + \rho$. This result is consistent with previous studies in flat space with $\eta_{0} = 2$ \cite{PhysRevB.27.5875}.

For exponential inflation ($d = 4$ and $\mu = 0$) we find a modification if $\rho \ge - 3$, i.e., the spatial noise correlator decreases at most like $\normsingle{\xbm}^{- 6}$. In general, the spectral index changes to
\begin{align}
	n_{\varphi}
		&=						d - 2 \eta_{0} - \rho,
								\label{eq:nvarphi-rho}
\end{align}
wherefore the choice $\rho = 0$ yields the short-range result.

The consequence is a dramatic change of the super-horizon power spectrum of the classical long-wavelength modes as compared to the case without noise. In particular, scale invariance does not hold on these scales.

Let us now return to our physical model of stochastic inflation. The split of the field $\vec{\varphi}$ into a long- and short-wavelength part, $\vec{\varphi} = \vec{\varphi}_{\t{\tiny L}} + \vec{\varphi}_{\t{\tiny S}}$, together with the free field equation, $( \Box + \mu^{2} )\vec{\varphi} = \vec{0}$, implies
\begin{align}
	\s( t, k )
		&=						\Bigg| \left( \Box_{k} + \mu^{2} \right) \bigg[ \Wrm_{\!\kappa}\!\left( \frac{ k }{ a( t ) } - \e \right) u( t, k ) \bigg] \Bigg|^{2} ,
								\label{eq:s(t,k)}
\end{align}
where $\Box_{k}$ is the (spatially Fourier transformed) covariant Laplacian, $u( t, k )$ is the mode function from equation \eqref{eq:varphi-fourier-decomposition}, and $\Wrm_{\!\kappa}$ is a smooth high-pass filter, cutting out the low frequencies below $\e$. The parameter $\kappa$ controls the width of the cut. In the limit $\kappa \ra 0$, $\Wrm_{\!\kappa}$ approaches a step function. Here we choose $\Wrm_{\!\kappa}( \cds ) = 1 / \pi \arctan( \cds / \kappa ) + 1 / 2$ and take $0 < \e \ll 1$ in order to cut far below the Hubble rate $H (= 1)$, and $\kappa \ll \e$ to have a narrow transition region between quantum and classical modes. We do not impose any restriction on $\mu$ except that we demand the radicant in \eqref{eq:ns} to be positive, i.e., $\mu^{2} \le 9 / 4$. 

It is natural to define a transition scale $k_{*}$ at which the two terms on the \rhs~of equation \eqref{eq:G=G0-sigmaG02} balance each other. It separates two regions such that for $k \gg k_{*}$ the behavior is noiseless and for $k \ll k_{*}$, dimensional reduction holds.

In the infra-red, the model of equation \eqref{eq:s(t,k)} is of long-range type. We find in four dimensions
\begin{align}
	\rho
		&=						3 \sqrt{ 1 - \frac{ 4 }{ 9 }\mu^2 } - 2,
\end{align}
and thus with \eqref{eq:nvarphi-rho} for $k \ll k_{*}$
\begin{align}
	n_{\varphi}
		&=						6 - 9 \sqrt{ 1 - \frac{ 4 }{ 9 }\mu^2 }
		=						- 3 + 2 \mu^{2} + \Ocal\!\left( \mu^{4} \right) ,
\end{align}
while for $k \gg k_{*}$ the noiseless spectral index \eqref{eq:ns} is recovered.

The zeros of the relative correction
\begin{align}
	\frac{ {\Grm_{\t{\tiny L}}}( t, k ) - {\Grm_{\t{\tiny L}}}_{0}( t, k ) }{ {\Grm_{\t{\tiny L}}}_{0}( t, k ) }
		&=						\s( t, k ) {\Grm_{\t{\tiny L}}}_{0}( t, k )
								\label{eq:deltaG}
\end{align}
define the transition scale $k_{*}( t )$. Its late-time behavior can be calculated analytically,
\begin{align}
	k_{*}( t )
		&\propto					\left( \frac{ \kappa^2 }{ \left( \e^2+\kappa^2\right)^2 } \right)^{\frac{1}{2 \sqrt{9-4 \mu^2}-2}}
								\left( \ee^{- t} \right)^{\frac{8-2 \sqrt{9-4 \mu^2}}{2 \sqrt{9-4 \mu^2}-2}} .
								\label{eq:kstar-exp-mu}
\end{align}
For $\mu = 0$ we find the asymptotic form
\begin{align}
	k_{*}( t )
		&=						\frac{ \ee^{- t / 2} \sqrt{\kappa}}{\sqrt{\pi }\sqrt{\e^2 + \kappa^2}}.
								\label{eq:kstar-exp-mu=0}
\end{align}
Thus, for $\e \ne 0$, $k_{*}$ goes to zero in the (step-function) limit $\kappa \ra 0$, i.e., dimensional reduction disappears. We should point out that the two limits $\kappa \ra 0$ and $\e \ra 0$ do not commute and that the case $\kappa \ne 0$ and $\e \ra 0$ is unphysical.

\begin{figure}[H]
	\centering
	\includegraphics[angle=0,scale=1]{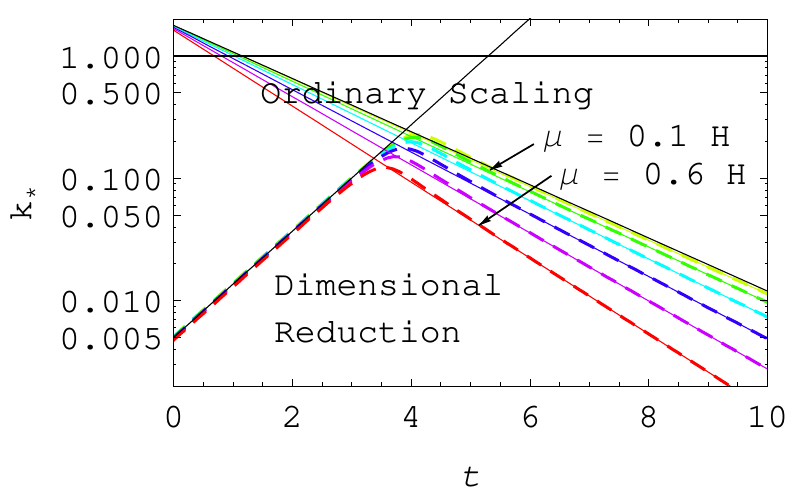}
	\caption{Comoving transition scale $k_{*}$ as a function of cosmic time $t$ (both in units of $H$) for mass $\mu / H = 0.1,\.0.2,\.0.3,\.0.4,\.0.5$ and $0.6\t{ (dashed 
			lines, top to bottom)}$. Dashed curves are numerical results, colored solid lines are analytic approximations, 
			and enveloping black lines are $\frac{ \e }{ 2 } a( t )$ and the asymptotic form \eqref{eq:kstar-exp-mu=0}, 
			respectively. A smoothing $\kappa = 10^{-3}$ and short-wavelength cut $\e = 10^{-2}$ are chosen.}
	\label{fig:kstar-exp}
\end{figure}

Figure \ref{fig:kstar-exp} shows the time behavior of the comoving scale $k_{*}$ for different values of the mass $\mu$. The solid rays represent the analytic approximation \eqref{eq:kstar-exp-mu}, while the dashed curves are obtained numerically from the roots of \eqref{eq:deltaG}. Well below this borderline the two-point function of the classical stochastic field obeys dimensional reduction, while well above ordinary scaling holds. One sees that $\mu = 0$ gives the largest $k_{*}( t )$. After an initial transient phenomenon, whose duration depends on the specific choice of $\e$ and $\kappa$, the comoving transition scale decays exponentially fast. Hence, the dimensional-reduction contribution is pushed to larger and larger scales as time increases. This therefore guarantees that quantum noise induces only a minor change of the spectral index on sub-horizon as well as on moderate super-horizon scales.

For concreteness, let us consider a mode with comoving $k = 0.05 H$. At time $t = 0$ it is within the region of ordinary scaling, suffering at most slightly from dimensional reduction. This mode enters then, after roughly two e-foldings, the region of broken scale invariance, but leaves it at the latest (for $\mu = 0$) after seven e-foldings and stays eternally in the scale-invariant regime, which itself grows exponentially fast.

One may connect the replica structure $\s$ to a non-linearity parameter $g_{\t{\tiny NL}}$, which shall now be defined via
\begin{align}
	\varphi^{}_{i}( t, k )
		&\equiv					\varphi^{\t{\tiny G}}_{i}( t, k ) - g_{\t{\tiny NL}}( t, k )\big( \varphi^{\t{\tiny G}}_{i}( t, k ) \big)^{2} ,
\end{align}
where $\vec{\varphi}^{\,\t{\tiny G}}( t, k )$ is a free Gaussian field. On the level of propagators, this translates to
\begin{align}
	{\Grm_{\t{\tiny L}}}( t, k )
		&=						{\Grm_{\t{\tiny L}}}_{0}( t, k ) + 3\,g_{\t{\tiny NL}}( t, k )^{2}\,{\Grm_{\t{\tiny L}}}_{0}( t, k )^{2}
\end{align}
and hence
\begin{align}
	\s( t, k )
		&=						3\,g_{\t{\tiny NL}}( t, k )^{2}
								\label{eq:sigma=13fNL2}
\end{align}
can be directly read off, using equation \eqref{eq:G=G0-sigmaG02}. $g_{\t{\tiny NL}}$ measures the influence of the quantum fluctuations, picked up by a smooth filter function. Formally, it resembles an effective non-Gaussianity parameter \cite{komatsu-2001-63} for the long-wavelength modes. However, this association is misleading since the theory we work with is Gaussian (but with non-trivial replica structure).

\begin{figure}
	\centering
	\includegraphics[angle=0,scale=1]{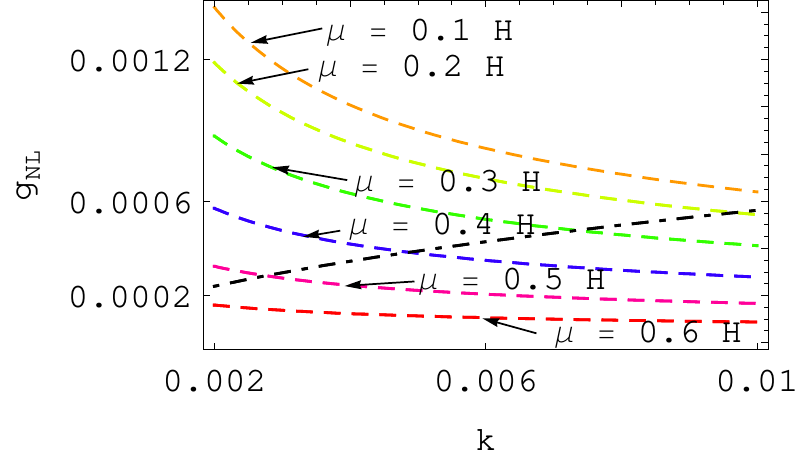}
	\caption{Non-linearity parameter $g_{\t{\tiny NL}}$ as a function of comoving momentum $k$ (in units of $H$) for mass $\mu / H = 0.1 \t{ (lowermost)},\.0.2,\.0.3,\.0.4,\.0.5$ 
			and $0.6 \t{ (uppermost)}$ with $H t = 10$, $\kappa = 10^{-3}$, and $\e = 10^{-2}$. The dotdashed black line represents the value of $k_{*}$ corresponding to the respective mass.}
	\label{fig:fNL-exponential}
\end{figure}

Figure \ref{fig:fNL-exponential} shows the dependence of $g_{\t{\tiny NL}}$ on the comoving momentum $k$ for various values of $\mu$ for fixed time $t = 10 H$, using equation \eqref{eq:sigma=13fNL2}. Firstly, one sees that increasing $\mu$ lifts the curve upwards, and secondly, one observers a divergence in the infra-red---displaying the effect of dimensional reduction. For $k \gg k_{*}$ one obtains a scale-invariant spectrum.

To summarize, we presented novel methods to study the power-spectrum of classical, stochastic fields in stochastic inflation. We demonstrate that replica field theory allows us to study the spatial behavior of non-coincident long-wavelength correlation functions. Dimensional reduction changes the spectral index on super-horizon scales, in the sense that it heavily amplifies the power spectrum of the classical modes in the infra-red. This effect has been calculated by a variational method, which allows us to go beyond ordinary perturbation theory and suggests it might be entirely of non-perturbative nature. However, in the limit of vanishing $\kappa$, i.e., of a sharp cut between long- and short-wavelength modes, the singularity disappears.

Those huge differences in the energy density on large super-horizon scales signal a breakdown of ordinary perturbation theory, because in that case one cannot speak about the spectrum of fluctuations in the usual, perturbative sense. It further displays the failure of the test-field assumption, since in the situation at hand it is no longer valid to neglect the back-reaction of the field on the geometry. However, the time-evolution of the long-wavelength field pushes the dimensionally reduced region exponentially fast to unobservable scales. This provides further support for the self-consistency of the idea of inflation, since regions of broken scale-invariance, with extraordinarily large fluctuations disappear faster than any causal patch of the universe expands.

It will be interesting to discuss other background space-times, as well as to include self-interactions. The present formalism might also help to tackle the full problem of stochastic inflation, where the geometry is also random. If we could assume that the dimensional reduction of our test field also applies to the inflaton, we would find further support of the idea of eternal inflation \cite{PhysRevD.27.2848}.

It is a pleasure to thank Benjamin Jurke, Daniel Kruppke, J{\'e}r{\^o}me Martin, Aravind Natarajan, Erandy Ramirez, Alexei Starobinsky, and Richard Woodard for stimulating discussions. FK acknowledges support from the Deutsche Forschungsgemeinschaft (DFG) under grant GRK 881.


\begin{thebibliography}{22}
\expandafter\ifx\csname natexlab\endcsname\relax\def\natexlab#1{#1}\fi
\expandafter\ifx\csname bibnamefont\endcsname\relax
  \def\bibnamefont#1{#1}\fi
\expandafter\ifx\csname bibfnamefont\endcsname\relax
  \def\bibfnamefont#1{#1}\fi
\expandafter\ifx\csname citenamefont\endcsname\relax
  \def\citenamefont#1{#1}\fi
\expandafter\ifx\csname url\endcsname\relax
  \def\url#1{\texttt{#1}}\fi
\expandafter\ifx\csname urlprefix\endcsname\relax\def\urlprefix{URL }\fi
\providecommand{\bibinfo}[2]{#2}
\providecommand{\eprint}[2][]{\url{#2}}

\bibitem[{\citenamefont{Starobinsky}(1982)}]{1982PhLB..117..175S}
\bibinfo{author}{\bibfnamefont{A.~A.} \bibnamefont{Starobinsky}},
  \bibinfo{journal}{Phys. Lett. B} \textbf{\bibinfo{volume}{117}},
  \bibinfo{pages}{175} (\bibinfo{year}{1982}).

\bibitem[{\citenamefont{Starobinsky}(1986)}]{1986LNP...246..107S}
\bibinfo{author}{\bibfnamefont{A.~A.} \bibnamefont{Starobinsky}}, in
  \emph{\bibinfo{booktitle}{Field Theory, Quantum Gravity and Strings}}
  (\bibinfo{year}{1986}), vol. \bibinfo{volume}{246} of
  \emph{\bibinfo{series}{Lecture Notes in Physics, Berlin, Springer}}, p.
  \bibinfo{pages}{107};
\bibinfo{author}{\bibfnamefont{S.-J.} \bibnamefont{{Rey}}},
  \bibinfo{journal}{Nucl. Phys. B} \textbf{\bibinfo{volume}{284}},
  \bibinfo{pages}{706} (\bibinfo{year}{1987});
\bibinfo{author}{\bibfnamefont{H.~E.} \bibnamefont{{Kandrup}}},
  \bibinfo{journal}{Phys. Rev. B} \textbf{\bibinfo{volume}{39}},
  \bibinfo{pages}{2245} (\bibinfo{year}{1989});
\bibinfo{author}{\bibfnamefont{D.~S.} \bibnamefont{Salopek}} \bibnamefont{and}
  \bibinfo{author}{\bibfnamefont{J.~R.} \bibnamefont{Bond}},
  \bibinfo{journal}{Phys. Rev. D} \textbf{\bibinfo{volume}{43}},
  \bibinfo{pages}{1005} (\bibinfo{year}{1991});
\bibinfo{author}{\bibfnamefont{A.~A.} \bibnamefont{Starobinsky}}
  \bibnamefont{and} \bibinfo{author}{\bibfnamefont{J.}~\bibnamefont{Yokoyama}},
  \bibinfo{journal}{Phys. Rev. D} \textbf{\bibinfo{volume}{50}},
  \bibinfo{pages}{6357} (\bibinfo{year}{1994}).

\bibitem[{\citenamefont{Vilenkin}(1983)}]{PhysRevD.27.2848}
\bibinfo{author}{\bibfnamefont{A.}~\bibnamefont{Vilenkin}},
  \bibinfo{journal}{Phys. Rev. D} \textbf{\bibinfo{volume}{27}},
  \bibinfo{pages}{2848} (\bibinfo{year}{1983});
\bibinfo{author}{\bibfnamefont{A.~S.}~\bibnamefont{{Goncharov}}},
  \bibinfo{author}{\bibfnamefont{A.~D.}~\bibnamefont{{Linde}}} \bibnamefont{and} \bibinfo{author}
  {\bibfnamefont{V.~F.}~\bibnamefont{{Mukhanov}}}, \bibinfo{journal}{Int.~J.~Mod.~Phys.~A}
  \textbf{\bibinfo{volume}{2}}, \bibinfo{pages}{561} (\bibinfo{year}{1987});
\bibinfo{author}{\bibfnamefont{A.}~\bibnamefont{Linde}},
  \bibinfo{author}{\bibfnamefont{D.}~\bibnamefont{Linde}}, \bibnamefont{and}
  \bibinfo{author}{\bibfnamefont{A.}~\bibnamefont{Mezhlumian}},
  \bibinfo{journal}{Phys. Rev. D} \textbf{\bibinfo{volume}{49}},
  \bibinfo{pages}{1783} (\bibinfo{year}{1994}).

\bibitem[{\citenamefont{Tsamis and Woodard}(2005)}]{Tsamis:2005hd}
\bibinfo{author}{\bibfnamefont{N.~C.} \bibnamefont{Tsamis}} \bibnamefont{and}
  \bibinfo{author}{\bibfnamefont{R.~P.} \bibnamefont{Woodard}},
  \bibinfo{journal}{Nucl. Phys. B} \textbf{\bibinfo{volume}{724}},
  \bibinfo{pages}{295} (\bibinfo{year}{2005}).

\bibitem[{\citenamefont{Liguori et~al.}(2004)\citenamefont{Liguori, Matarrese,
  Musso, and Riotto}}]{liguori-2004-0408}
\bibinfo{author}{\bibfnamefont{M.}~\bibnamefont{Liguori}},
  \bibinfo{author}{\bibfnamefont{S.}~\bibnamefont{Matarrese}},
  \bibinfo{author}{\bibfnamefont{M.}~\bibnamefont{Musso}}, \bibnamefont{and}
  \bibinfo{author}{\bibfnamefont{A.}~\bibnamefont{Riotto}},
  \bibinfo{journal}{JCAP} \textbf{\bibinfo{volume}{0408}}, \bibinfo{pages}{011}
  (\bibinfo{year}{2004}).

\bibitem[{\citenamefont{{M{\'e}zard} and {Parisi}}(1991)}]{1991JPhy1...1..809M}
\bibinfo{author}{\bibfnamefont{M.}~\bibnamefont{{M{\'e}zard}}}
  \bibnamefont{and} \bibinfo{author}{\bibfnamefont{G.}~\bibnamefont{{Parisi}}},
  \bibinfo{journal}{J. Phys. I} \textbf{\bibinfo{volume}{1}},
  \bibinfo{pages}{809} (\bibinfo{year}{1991}).

\bibitem[{\citenamefont{Leach et~al.}(2002)\citenamefont{Leach, Liddle, Martin,
  and Schwarz}}]{PhysRevD.66.023515}
\bibinfo{author}{\bibfnamefont{S.~M.} \bibnamefont{Leach}},
  \bibinfo{author}{\bibfnamefont{A.~R.} \bibnamefont{Liddle}},
  \bibinfo{author}{\bibfnamefont{J.}~\bibnamefont{Martin}}, \bibnamefont{and}
  \bibinfo{author}{\bibfnamefont{D.~J.} \bibnamefont{Schwarz}},
  \bibinfo{journal}{Phys. Rev. D} \textbf{\bibinfo{volume}{66}},
  \bibinfo{pages}{023515} (\bibinfo{year}{2002}).

\bibitem[{\citenamefont{{M{\'e}zard} et~al.}(1988)\citenamefont{{M{\'e}zard},
  {Parisi}, {Virasoro}, and {Thouless}}}]{1988PhT....41l.109M}
\bibinfo{author}{\bibfnamefont{M.}~\bibnamefont{{M{\'e}zard}}},
  \bibinfo{author}{\bibfnamefont{G.}~\bibnamefont{{Parisi}}},
  \bibinfo{author}{\bibfnamefont{M.~A.} \bibnamefont{{Virasoro}}},
  \bibnamefont{and} \bibinfo{author}{\bibfnamefont{D.~J.}
  \bibnamefont{{Thouless}}}, \bibinfo{journal}{Physics Today}
  \textbf{\bibinfo{volume}{41}}, \bibinfo{pages}{109} (\bibinfo{year}{1988}).

\bibitem[{\citenamefont{Fedorenko and K{\"u}hnel}(2007)}]{fedorenko-2007}
\bibinfo{author}{\bibfnamefont{A.~A.} \bibnamefont{Fedorenko}}
  \bibnamefont{and}
  \bibinfo{author}{\bibfnamefont{F.}~\bibnamefont{K{\"u}hnel}},
  \bibinfo{journal}{Phys. Rev. B} \textbf{\bibinfo{volume}{75}},
  \bibinfo{pages}{174206} (\bibinfo{year}{2007}).

\bibitem[{\citenamefont{Feynman}(1955)}]{PhysRev.97.660}
\bibinfo{author}{\bibfnamefont{R.~P.} \bibnamefont{Feynman}},
  \bibinfo{journal}{Phys. Rev.} \textbf{\bibinfo{volume}{97}},
  \bibinfo{pages}{660} (\bibinfo{year}{1955}).

\bibitem[{\citenamefont{Jensen}(1906)}]{Jensen-1906-1}
\bibinfo{author}{\bibfnamefont{J.~L.~W.~V.}~\bibnamefont{Jensen}},
  \bibinfo{journal}{Acta Math.} \textbf{\bibinfo{volume}{30}},
  \bibinfo{pages}{175} (\bibinfo{year}{1906}).

\bibitem[{\citenamefont{Kardar et~al.}(1983)\citenamefont{Kardar, McClain, and
  Taylor}}]{PhysRevB.27.5875}
\bibinfo{author}{\bibfnamefont{M.}~\bibnamefont{Kardar}},
  \bibinfo{author}{\bibfnamefont{B.}~\bibnamefont{McClain}}, \bibnamefont{and}
  \bibinfo{author}{\bibfnamefont{C.}~\bibnamefont{Taylor}},
  \bibinfo{journal}{Phys. Rev. B} \textbf{\bibinfo{volume}{27}},
  \bibinfo{pages}{5875} (\bibinfo{year}{1983}).

\bibitem[{\citenamefont{Parisi and Sourlas}(1979)}]{PhysRevLett.43.744}
\bibinfo{author}{\bibfnamefont{G.}~\bibnamefont{Parisi}} \bibnamefont{and}
  \bibinfo{author}{\bibfnamefont{N.}~\bibnamefont{Sourlas}},
  \bibinfo{journal}{Phys. Rev. Lett.} \textbf{\bibinfo{volume}{43}},
  \bibinfo{pages}{744} (\bibinfo{year}{1979});
\bibinfo{author}{\bibfnamefont{A.}~\bibnamefont{Aharony}},
  \bibinfo{author}{\bibfnamefont{Y.}~\bibnamefont{Imry}}, \bibnamefont{and}
  \bibinfo{author}{\bibfnamefont{S.}~\bibnamefont{Ma}}, \bibinfo{journal}{Phys.
  Rev. Lett.} \textbf{\bibinfo{volume}{37}}, \bibinfo{pages}{1364}
  (\bibinfo{year}{1976}).

\bibitem[{\citenamefont{{Klein} et~al.}(1984)\citenamefont{{Klein}, {Landau},
  and {Perez}}}]{1984CMaPh..94..459K}
\bibinfo{author}{\bibfnamefont{A.}~\bibnamefont{{Klein}}},
  \bibinfo{author}{\bibfnamefont{L.~J.} \bibnamefont{{Landau}}},
  \bibnamefont{and} \bibinfo{author}{\bibfnamefont{J.~F.}
  \bibnamefont{{Perez}}}, \bibinfo{journal}{Comm. Math. Phys.}
  \textbf{\bibinfo{volume}{94}}, \bibinfo{pages}{459} (\bibinfo{year}{1984}).

\bibitem[{\citenamefont{Komatsu and Spergel}(2001)}]{komatsu-2001-63}
\bibinfo{author}{\bibfnamefont{E.}~\bibnamefont{Komatsu}} \bibnamefont{and}
  \bibinfo{author}{\bibfnamefont{D.~N.} \bibnamefont{Spergel}},
  \bibinfo{journal}{Phys. Rev. D} \textbf{\bibinfo{volume}{63}},
  \bibinfo{pages}{063002} (\bibinfo{year}{2001}).

\end{thebibliography}
\end{document}